\documentclass[amsmath,longbibliography,superscriptaddress,reprint]
{revtex4-1}
\usepackage{hyperref}
\usepackage{graphicx}
\begin{document}
\title{Near-unity radiative quantum efficiency of excitons in carbon nanotubes}
\author{H.~Machiya}
\affiliation{Nanoscale Quantum Photonics Laboratory, RIKEN Cluster for Pioneering Research, Saitama, 351-0198, Japan}
\affiliation{Department of Electrical Engineering, The University of Tokyo, Tokyo, 113-8656, Japan}
\author{D.~Yamashita}
\affiliation{Quantum Optoelectronics Research Team, RIKEN Center for Advanced Photonics, Saitama, 351-0198, Japan}
\author{A.~Ishii}
\affiliation{Nanoscale Quantum Photonics Laboratory, RIKEN Cluster for Pioneering Research, Saitama, 351-0198, Japan}
\affiliation{Quantum Optoelectronics Research Team, RIKEN Center for Advanced Photonics, Saitama, 351-0198, Japan}
\author{Y.~K.~Kato}
\email[Corresponding author: ]{yuichiro.kato@riken.jp}
\affiliation{Nanoscale Quantum Photonics Laboratory, RIKEN Cluster for Pioneering Research, Saitama, 351-0198, Japan}
\affiliation{Quantum Optoelectronics Research Team, RIKEN Center for Advanced Photonics, Saitama, 351-0198, Japan}
\begin{abstract}
The efficiencies of photonic devices are primarily governed by radiative quantum efficiency, which is a property given by the light emitting material. Quantitative characterization for carbon nanotubes, however, has been difficult despite being a prominent material for nanoscale photonics.
Here we determine the radiative quantum efficiency of bright excitons in carbon nanotubes by modifying the exciton dynamics through cavity quantum electrodynamical effects.
Silicon photonic crystal nanobeam cavities are used to induce the Purcell effect on individual carbon nanotubes.
Spectral and temporal behavior of the cavity enhancement is characterized by photoluminescence microscopy, and the fraction of the radiative decay process is evaluated.
We find that the radiative quantum efficiency is near unity for bright excitons in carbon nanotubes at room temperature.
\end{abstract}

\maketitle

Radiative quantum efficiency is a fundamental physical quantity that ultimately limits the efficiency of optoelectronic devices.
III-V direct gap semiconductors with radiative quantum efficiencies close to 100\% have played a central role in the development of efficient light emitting diodes (LEDs) that are now used ubiquitously \cite{Schubert2005}.
Modern high-power diode lasers see extensive use in industries including telecommunications \cite{Milonni2010}, materials processing, and medical treatments \cite{Steen2010}, but the performance needed for broad applications would have not been possible without the high quantum efficiencies of the gain materials.
Organic semiconductors have likewise seen commercial success in LED displays \cite{Chen2018}, but only after substantial improvement of their radiative quantum efficiencies \cite{Adachi2001,Lamansky2001}. 

Carbon nanotubes (CNTs), also as a direct gap semiconductor, have demonstrated their potential as a key material for nanoscale photonic devices.
Electrically-gated $pn$-junction devices can be constructed from nanotubes with diameters significantly smaller than the wavelength, where extremely efficient photocurrent generation \cite{Gabor:2009,McCulley2020} and excitonic electroluminescence \cite{Higashide2017,Graf2017} have been demonstrated.
Unique exciton physics can be exploited to generate telecom band single photons at room temperature \cite{Ma2015,Ishii2017,He2018}, and the compatibility of nanotube emitters with silicon photonics \cite{Miura2014,Pyatkov2016,Ishii2018,Machiya2018,Noury2015,Higuchi2020} offers opportunities in integrated quantum optics \cite{Khasminskaya2016}.
The device performance is again limited by the radiative quantum efficiency, but uncertainty of absorption cross section and sensitivity to nanotube quality have made quantitative characterization difficult.

Here we experimentally determine the radiative quantum efficiencies of excitons in air-suspended CNTs. Quantum electrodynamical effects in nanoscale photonic cavities are used to selectively modify the radiative decay rate of excitons, allowing us to gain insight to the fractions of radiative and nonradiative processes. Individual CNTs are coupled to air-mode photonic crystal nanobeam cavities, and photoluminescence (PL) measurements are performed to quantitatively evaluate the spectral and temporal enhancements induced by the Purcell effect. We find the radiative quantum efficiencies of bright excitons to be near unity at room temperature.

\section*{Results}

\subsection*{Quantum electrodynamical modification of radiative decay rates}

We start by discussing how microcavities alter the decay dynamics of excitons through a quantum electrodynamical phenomenon known as the Purcell effect \cite{Purcell1946}. In free space, the total exciton decay rate can be written as  $\gamma_{\text{r}}+\gamma_{\text{nr}}$ where $\gamma_{\text{r}}$ and $\gamma_{\text{nr}}$ are the radiative and non-radiative decay rates, respectively.
When excitons are coupled to a microcavity, the increased photon density of states gives rise to an additional radiative decay rate $F\gamma_{\text{r}}$, where $F$ is the Purcell factor \cite{Miura2014,Pyatkov2016,Ishii2018,Machiya2018,Jeantet2016,Luo2017}. The accelerated radiative recombination appears in the spectral domain as an emission enhancement at the cavity mode frequency, whereas it is more directly observed in the time domain as a shortening in the lifetime.
The overall change in the decay rate can be characterized by the acceleration factor
\begin{equation}
\label{Eq1}
A=\frac{(F+1)\gamma_{\text{r}}+\gamma_{\text{nr}}}{\gamma_{\text{r}}+\gamma_{\text{nr}}}=1+F\eta,
\end{equation}
where $\eta=\gamma_{\text{r}}/(\gamma_{\text{r}}+\gamma_{\text{nr}})$ is the radiative quantum efficiency.
This relationship allows us to determine the quantum efficiency if $F$ and $A$ are known.
By performing spectral- and time-domain measurements on excitons coupled to a microcavity, we evaluate these values independently.

\begin{figure}
\includegraphics{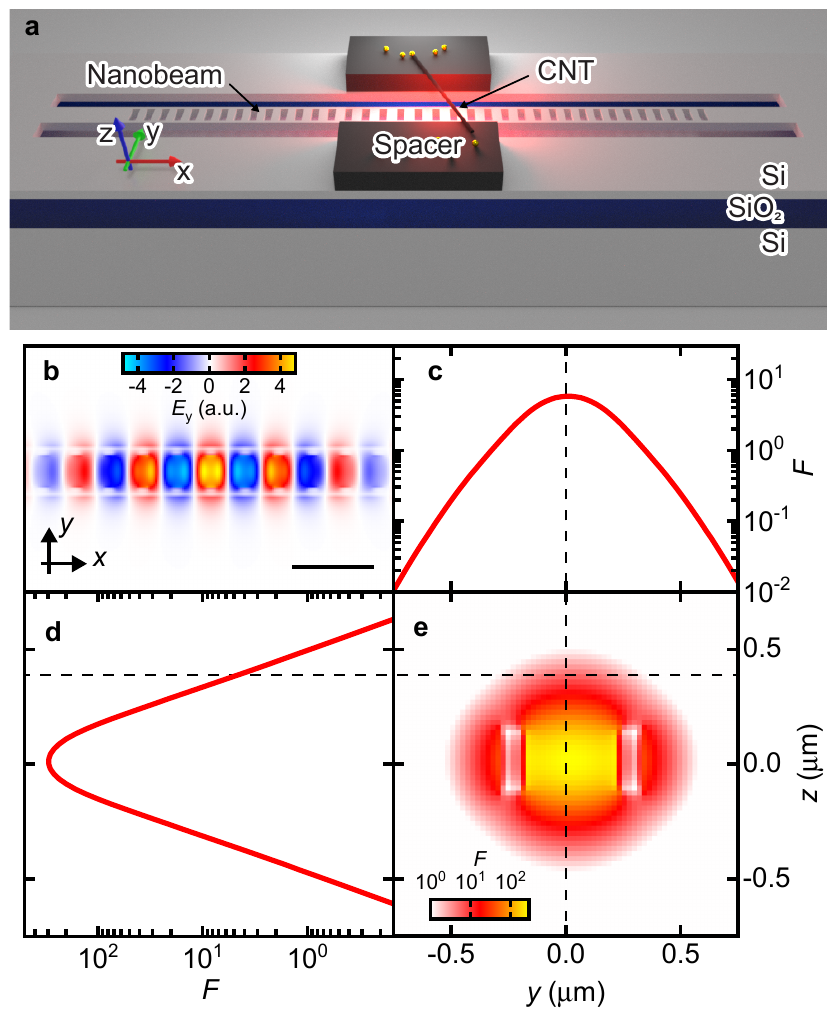}
\caption{\label{Fig1}(a) Schematic of the device. (b) Simulated spatial distribution of the $y$-component of the electric field $E_y$ for the fundamental transverse-electric mode at $z=0.00~\mu$m. The scale bar is 1~$\mu$m. (c-e) Spatial profiles of the calculated Purcell factor at $x=0.00~\mu$m. Cross sections through the dashed lines at (c) $z=0.38~\mu$m and (d) $y=0.00~\mu$m. For (b-e), The coordinate origin is taken to be the center of the cavity.}
\end{figure}

As our primary interest is in the intrinsic properties of excitons, we use as-grown air-suspended CNTs for the determination of the quantum efficiency.
These tubes are known to exhibit high PL yields, indicative of their pristine nature \cite{Lefebvre:2006}. To modify the radiative decay rates by the Purcell effect, the air-suspended tubes are coupled to silicon photonic crystal nanobeam cavities \cite{Miura2014,Machiya2018} as shown in Fig.~\ref{Fig1}a.
We utilize air-mode cavities characterized by ultrasmall mode volumes and large electric fields in the air holes, which allows for high efficiency coupling \cite{Miura2014}.  
The cavities are designed by finite-difference time-domain (FDTD) simulations for sufficiently high Purcell factors.
The mode profile is presented in Fig.~\ref{Fig1}b, where the large fields in the air holes can be seen. The cavity has a resonant wavelength $\lambda=1460$~nm, and its quality factor and mode volume are calculated to be $4\times10^5$ and $0.025\times\lambda^3$, respectively.
We note that the Purcell factor is insensitive to the cavity quality factor as it is sufficiently high compared to the emitter quality factor of $\sim$100 \cite{Exter1996}.

Because CNTs suffer from strong quenching effect upon contact with the substrate \cite{Sarpkaya:2013,Ishii2015}, we take special care to suspend the tubes a few hundred nanometers above the cavity by introducing spacer layers.
The thickness of the spacers is critical in our device design since the evanescent fields of the cavity mode decreases exponentially above the cavity.
Figures~\ref{Fig1}c-e illustrate the position dependence of the calculated Purcell factor.
Due to the ultrasmall mode volume of the cavity, a Purcell factor much larger than unity is readily available even a few hundred nm above the cavity surface (Fig.~\ref{Fig1}d dashed line).

\subsection*{Device fabrication and spectroscopic characterization}

\begin{figure}
\includegraphics{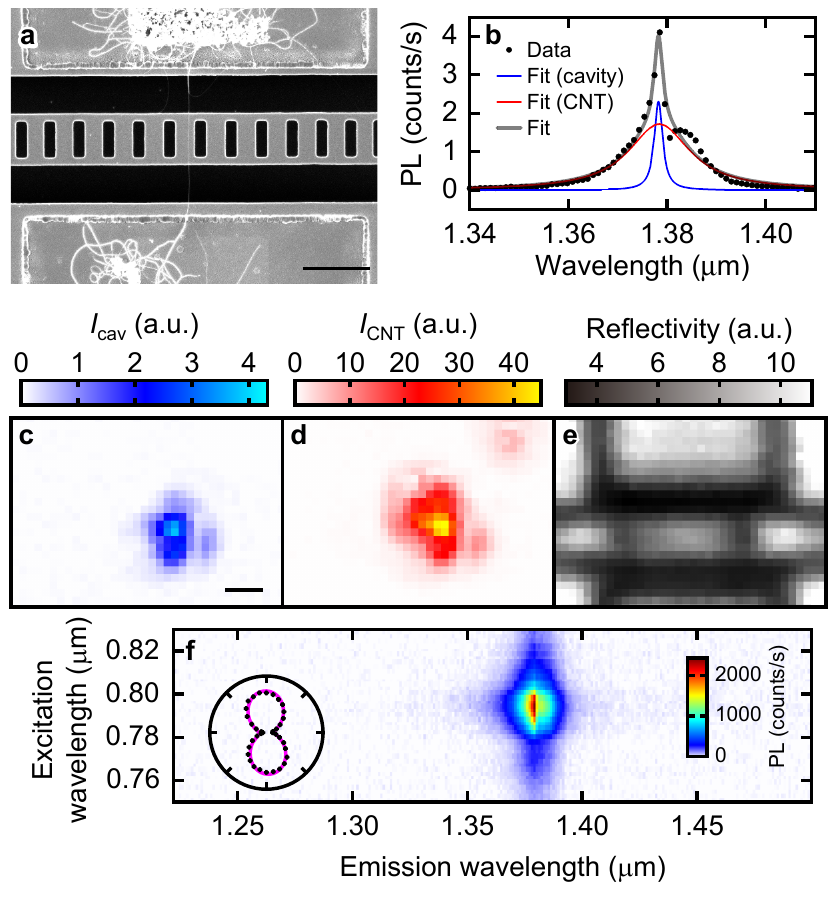}
\caption{\label{Fig2}(a) Scanning electron micrograph of a fabricated device. A CNT is suspended across the width of the trench near the center of the image. The scale bar is 1~$\mu$m. (b) Typical PL spectrum of a device showing optical coupling to the cavity. The dots are data and the gray line is the bi-Lorentzian fit. The blue and red curves correspond to the cavity and the CNT peak components, respectively. The cavity peak is centered at 1378.1 nm and has a linewidth of 0.5 nm, whereas the CNT peak has a center wavelength of 1378.4 nm with a linewidth of 14.1 nm. (c,d) PL images for the (c) cavity and (d) CNT emission, obtained by mapping out the peak areas for each peak. (e) Reflectivity image taken simultaneously with (c) and (d). The scale bar in (c) is 1~$\mu$m and is shared between panels (c-e). For (b-e), the excitation laser wavelength is 793~nm and the laser polarization is parallel to the nanotube axis. (f) PL excitation map for the same device, taken with the laser polarization perpendicular to the nanobeam axis. Chirality is assigned to (9,8) using tabulated data \cite{Ishii2015}. Inset: excitation polarization dependence of PL at 1378~nm with a spectral integration window of 5~nm. The dots are data and the line is a fit by cosine squared. (b-f) are taken with an excitation power of 20~$\mu$W.}
\end{figure}

Based on the simulation results, we choose a spacer layer thickness of 250~nm for our devices and fabricate the cavities from a silicon-on-insulator wafer.
Catalyst is placed on the spacer layers beside the cavity, and CNTs are directly synthesized over the cavities by chemical vapor deposition \cite{Ishii2015}.
A scanning electron micrograph of a device is shown in Fig.~\ref{Fig2}a. 
As the chirality and the location of the nanotubes are randomly distributed, some effort is required to find tubes that have good spatial and spectral overlap with the cavity modes. 
To overcome the low yield of devices with optical coupling which is typically less than 0.1\%, more than 300,000 devices in total are prepared. Automated scanning is performed to efficiently collect PL spectra from a large number of devices, and we identify spectral signatures of the cavity mode to select devices for detailed investigation. 

Figure~\ref{Fig2}b shows a PL spectrum of a representative device, where a sharp peak corresponding to the cavity mode can be seen on top of a broader peak from CNT emission into free space.
The cavity peak area $I_{\text{cav}}$ and the CNT peak area $I_{\text{CNT}}$ can be extracted using bi-Lorentzian fitting, and a two-dimensional PL scan is performed to map out the spatial extent of the two components (Fig.~\ref{Fig2}c-d).
A reflectivity image is simultaneously taken and is shown in Fig.~\ref{Fig2}e.
The profile of the cavity emission is predominantly localized at the center of the device, as expected from the electric field distribution of the cavity mode.
In comparison, the nanotube peak component extends beyond the width of the nanobeam, showing emission from the full length of the suspended nanotube.
We further characterize the device to ensure that the tube is individual and clean.
A PL excitation map is used to identify the chirality, whereas the angle of the nanotube is determined from polarization dependence of PL (Fig.~\ref{Fig2}f).
Devices showing multiple peaks or temporal instabilities are eliminated from further measurements, as they indicate bundling or contamination \cite{Tan:2007,Matsuda2005a}.

The Purcell factor is extracted from the PL spectrum taken at the center of the device (Fig.~\ref{Fig2}b). The ratio of the cavity accelerated radiative decay rate to the free-space radiative decay rate gives $F$, which can be evaluated using the PL intensities of the two peak components.
Letting $C_{\text{cav}}$ and $C_{\text{CNT}}$ be the collection efficiencies of the cavity mode and the nanotube emission, respectively,

\begin{equation}
F=\frac{I_{\text{cav}}/C_{\text{cav}}}{I_{\text{CNT}}/C_{\text{CNT}}}.
\end{equation}

\begin{figure}
\includegraphics{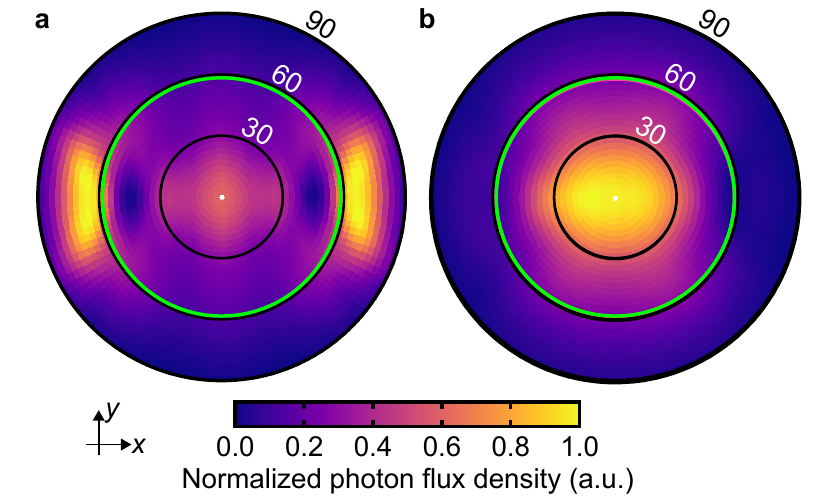}
\caption{\label{Fig3}Calculated far-field radiation pattern of (a) the fundamental cavity mode and (b) the uncoupled nanotube emission. Photon flux density is plotted as a function of polar and azimuthal angles in spherical coordinates. The radial axis represents the polar angle. The green circle represents the numerical aperture of the objective lens. }
\end{figure}

The collection efficiencies are obtained using the radiation patterns from FDTD simulations.
Since the cavity mode primarily consists of zone boundary waveguide modes, coupling to leaky modes above the light line is small \cite{Quan:2011}.
Most of the light that leaks out is directed at a low angle along the nanobeam (Fig.~\ref{Fig3}a), likely because most of the coupling occurs near the light line.
As a result, only a small fraction of the cavity photons fit within the numerical aperture of the objective lens, giving $C_{\text{cav}}=0.15$.
In contrast, nanotube radiation is reduced by the photonic bandgap at low angles, causing the emission to be redirected upward and increasing the collection efficiency to 40\% (Fig.~\ref{Fig3}b).
Also taking into account the light collected from the length of the tube that extends beyond the nanobeam, we obtain $C_{\text{CNT}}=1.6$.
This analysis results in Purcell factor $F=0.84\pm0.22$ for this device, a reasonable value considering the lateral displacement of $\sim0.6~\mu$m (Fig.~\ref{Fig2}c-e) from  the cavity field maxima and the simulation in Fig.~\ref{Fig1}c-e.

\subsection*{Time-resolved measurements}

\begin{figure}
\includegraphics{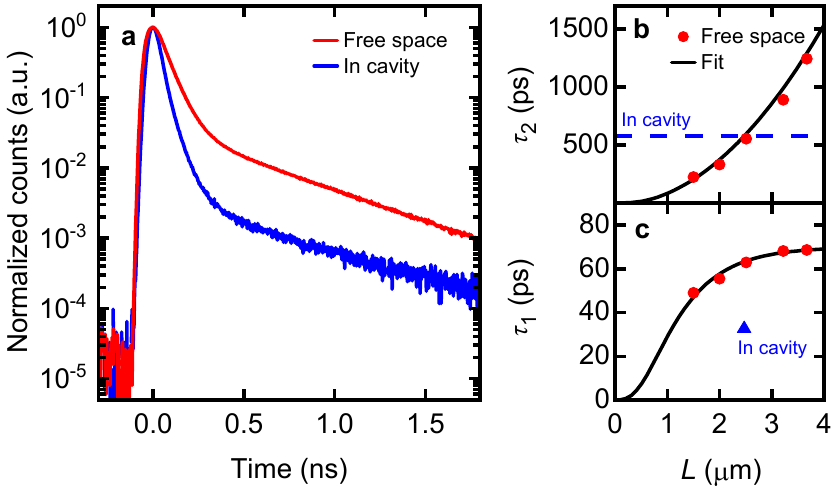}
\caption{\label{Fig4}(a) Decay curves for the Purcell-enhanced nanotube emission from the same device as in Fig.~\ref{Fig2} (blue) and a 2.0-$\mu$m-long (9,8) nanotube in free space (red). The excitation powers are 100~nW and 5~nW for nanotubes in the cavity and in free space, respectively. (b,c) Length dependence of (b) $\tau_2$ and (c) $\tau_1$ in free space. The red dots are data and the lines are fits using the exciton diffusion model \cite{Ishii2019}. The blue dashed line in (b) shows the dark exciton lifetime in the cavity, from which we extract the effective nanotube length. The blue triangle in (c) shows the Purcell-accelerated bright exciton lifetime.}
\end{figure}

In order to determine the quantum efficiency, we now need to evaluate the acceleration factor by performing time-resolved PL measurements. 
The same PL microscopy setup is used but the devices are excited with $\sim$100~fs laser pulses and emission is detected by a fiber-coupled superconducting single photon detector.
Time-resolved PL data taken from the device shown in Fig.~\ref{Fig2} is plotted as a blue curve in Fig.~\ref{Fig4}a.
We observe fast and slow decay components with lifetimes $\tau_1$ and $\tau_2$, respectively.
By comparing the decay curve to that of a nanotube in free space having a similar length (red curve in Fig.~\ref{Fig4}a), we already see that $\tau_1$ corresponding to the bright exciton lifetime is considerably shorter, while $\tau_2$ reflecting the dark exciton dynamics is comparable.
This is expected as the bright exciton decay is accelerated by the Purcell effect, whereas the dark excitons do not interact with photons and therefore their lifetime would not be changed by the cavity.
We note that the contrasting behavior of bright and dark exciton lifetimes rules out extrinsic quenching effects as the cause for bright exciton lifetime shortening.
Such effects including quenching caused by contact with the substrate should shorten both bright and dark exciton lifetimes, as in the case for end quenching \cite{Ishii2019}.

The bright exciton lifetime is extracted by fitting the data for the device with a biexponential function, and we obtain $\tau_1=34.8\pm0.3$~ps and $\tau_2=586\pm5$~ps.
The change in $\tau_1$ for the nanotube coupled to the cavity compared with a nanotube in free space is needed to obtain the value of $A$, but care must be taken since $\tau_1$ is dependent on the nanotube length $L$.
We take advantage of the fact that the dark exciton dynamics is unaffected by the cavity, and use the length dependence of $\tau_2$ to determine the nanotube length at which $\tau_1$ is compared.

Figures~\ref{Fig4}b and \ref{Fig4}c show the length dependence of $\tau_2$ and $\tau_1$, respectively, for air-suspended (9,8) nanotubes in free space \cite{Ishii2019}.
Both the bright and dark exciton lifetimes increase with the nanotube length due to reduced end quenching in longer nanotubes.
By looking up the nanotube length for $\tau_2=586$~ps (Fig.~\ref{Fig4}b dashed line), $L=2.5\pm0.1~\mu$m is obtained.
The acceleration factor is then evaluated using the corresponding bright exciton lifetime in free space (Fig.~\ref{Fig4}c), and we find $A=1.82\pm0.09$.
Having determined both $F$ and $A$ for this particular device, the radiative quantum efficiency of bright excitons is calculated to be $0.98\pm0.11$ using Eq.~\ref{Eq1}.

\begin{figure}
\includegraphics{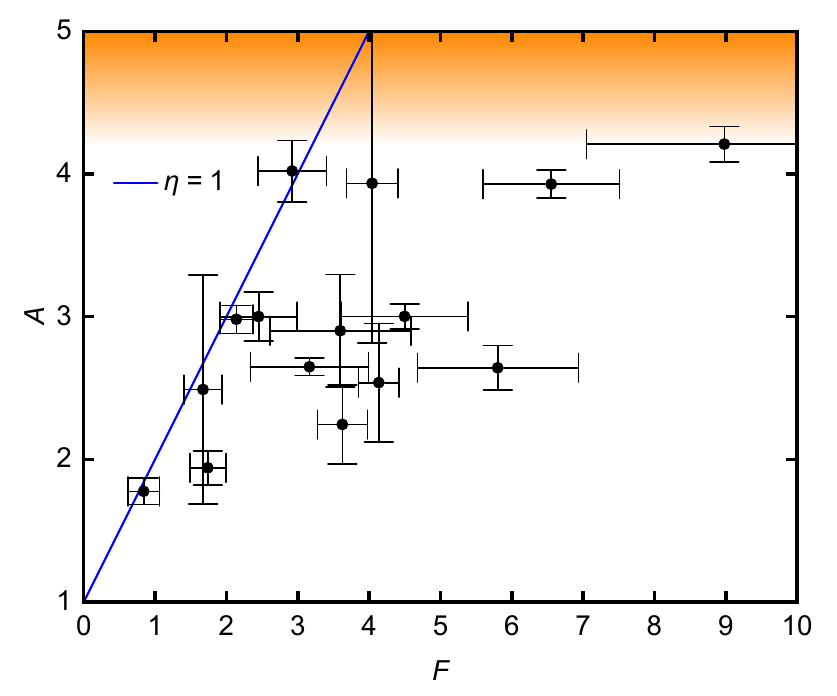}
\caption{\label{Fig5}The acceleration factor as a function of the Purcell factor. The dots are data and the error bars show the 1$\sigma$ confidence interval. The blue line indicates $\eta=1$ as given by Eq.~\ref{Eq1}. The orange shaded area corresponds to the region beyond the shortest resolvable lifetimes. Chiralities of the nanotubes in the measured devices are (9,7), (9,8), (10,8), and (11,6).}
\end{figure}

The near-unity quantum efficiency is reproducibly observed in a number of devices.
We have repeated the same measurements in additional devices with varying Purcell factors, and the results are summarized in Supplementary Table~1.
The acceleration factor is plotted as a function of the Purcell factor in Fig.~\ref{Fig5}, where variations and errors in the measured quantum efficiencies can be seen.
Data points closer to the top-left corner indicate higher quantum efficiency, and we observe that the blue line representing unity quantum efficiency passes through a number of data points within the error bars.
It is not surprising that there are devices showing lower efficiencies, as undetected defects or contamination can introduce nonradiative recombination.
We may therefore interpret the highest efficiencies observed as being the intrinsic property of the bright excitons in CNTs.
Even when considering the error bars of the data points, it is fair to state that their quantum efficiencies are near unity.

\section*{Discussion}
The radiative quantum efficiency is directly related to PL yield which is defined as a fraction of emitted photons to absorbed photons.
The PL yield is not necessarily equal to the radiative quantum efficiency of the bright excitons, as most of the absorbed photons end up in dark excitonic states.
There exist 16 exciton states since both the valence and the conduction bands are 4-fold degenerate, out of which only one state is bright and all the other states are dark due to spin, momentum, and parity selection rules \cite{Miyauchi2013a,Perebeinos:2005,Spataru:2005}.
All 16 states should be populated equally if free carriers are generated as in the case of $E_{22}$ excitation \cite{Kumamoto2014}, and a PL yield of 6.25\% would result for unity quantum efficiency of bright excitons.
Experimentally, a similar value of $\sim$7\% has been reported for air-suspended nanotubes at $E_{22}$ excitation \cite{Lefebvre:2006}.

It should be possible to enhance the PL yield by choosing excitation methods with higher initial population of bright excitons.
Resonant excitation can be used to selectively populate the bright states, and in principle unity PL yield would be achieved.
If rejection of excitation laser becomes an experimental issue, an alternative is to utilize phonon sidebands \cite{Torrens2008} and $E_{11}$ excited states \cite{Lefebvre2008}.
By limiting excitation into the spin singlet manifold, the PL yield is expected to improve by a factor of 4.
A more sophisticated scheme will be needed for electrical excitation, such as spin polarized \cite{Wang2012} or energy selective \cite{Kimura2019} injection.

Lastly, we would like to comment on the radiative lifetime of excitons in CNTs \cite{Perebeinos:2005,Spataru:2005,Miyauchi:2009}.
When radiative quantum efficiency is unity, bright exciton decay time is equivalent to the radiative lifetime.
We can therefore consider the reported bright exciton decay times of air-suspended nanotubes ranging from 60~ps to 90~ps to be their radiative lifetime \cite{Ishii2019}.
In comparison, theoretical calculations \cite{Perebeinos:2005,Spataru:2005} have yielded radiative lifetimes on the order of 10~ns for micelle-wrapped nanotubes, although the dielectric environment is considerably different from those in the air-suspended structures.
The experimentally observed bright exciton decay times also show a clear family pattern \cite{Ishii2019}, which has not been addressed theoretically.
Updated calculations are needed for an accurate description of radiative lifetimes in air-suspended nanotubes.

In conclusion, we have quantitatively determined the radiative quantum efficiency of bright excitons in air-suspended CNTs. The fraction of the radiative recombination is extracted by utilizing quantum electrodynamical effects in air-mode nanobeam cavities. Combining the Purcell factor characterized from the PL spectra and the acceleration factor obtained from the time-resolved measurements, we find that the radiative quantum efficiency of bright excitons is near unity at room temperature. Our results reveal the intrinsic property of bright excitons in CNTs and demonstrate their true potential for nanoscale photonics. With high-efficiency light emission and single-photon generation capabilities, CNTs may bring breakthroughs in subwavelength silicon photonics and integrated quantum optics.

\section*{Methods}

\paragraph*{Finite-difference time-domain simulations}

We use an open-source FDTD calculation library MEEP \cite{Oskooi2010} to simulate the cavity modes. The cavity consists of a periodic array of 39 rectangular holes in an air-bridge nanobeam structure, and the periodicity is increased at the cavity center in a parabolic manner over 16 periods \cite{Miura2014}. The lattice constant $a=350$~nm is used and the cavity center period is 1.18$a$. The beam width is 600~nm and the holes are 140~nm by 400~nm. The Si layer thickness of 260~nm is chosen to match the actual devices. Position dependent Purcell factor is obtained from $F(\vec{r})=\frac{3}{4\pi^2}\frac{Q}{V}\lambda^3|E_y(\vec{r})|^2/\max|E_y(\vec{r})|^2$, where $Q$ is the emitter quality factor, $V$ is the mode volume, $\lambda$ is the wavelength, and $E_y(\vec{r})$ is the $y$ component of the electric field amplitude at position $\vec{r}$.

\paragraph*{Device fabrication}

We fabricate the nanobeam cavities from a silicon-on-insulator wafer with a 1-$\mu$m-thick buried oxide layer and a 260-nm-thick top silicon layer.
The nanobeams are designed to be 800~nm wide and the hole width is fixed at 500~nm, where we have chosen the parameters to correct for fabrication errors.
The lattice constant $a$ is varied from 320~nm to 511~nm to tune the resonance, and the hole length is scaled to be $0.4a$.
We prepare 30,000 devices on a chip, covering a wide wavelength range from 1100 to 1600 nm for coupling to a variety of nanotube chiralities.
Cavities and alignment marks are patterned by electron beam lithography, and the top silicon layer is etched by an inductively-coupled plasma etcher.
We perform another electron beam writing to define the spacer regions, where Si is sputtered and subsequently lifted off.
The buried oxide layer is then etched by hydrofluoric acid to make the air-bridge structure.
We perform a final electron beam writing to define catalyst windows, and the completed chip is diced into 4~mm square chips.
Catalyst solution is prepared by dispersing Fe(III) acetylacetonate and dry fumed silica in ethanol by ultrasonication, and the catalyst particles are deposited by spin-coating and lift-off processes.
We finally synthesize the nanotubes by chemical vapor deposition at 800$^\circ$C for 1~minute using ethanol as a carbon source \cite{Ishii2015}.

\paragraph*{Photoluminescence microspectroscopy}
The devices are characterized in a home-built confocal microscope system \cite{Ishii2015,Machiya2018}. Samples are mounted on a motorized three-dimensional feedback stage, allowing us to perform automated measurements over a large number of devices.
A wavelength-tunable Ti:sapphire laser is used as an excitation source, and the beam is focused by an objective lens with a numerical aperture of 0.85 and a focal length of 1.8~mm.
Emission is collected with the same objective lens, and coupled to a grating spectrometer with a liquid-nitrogen-cooled InGaAs photodiode array through a confocal pinhole corresponding to an aperture with 2.7~$\mu$m diameter at the sample image plane.
Laser reflection is simultaneously collected and is monitored by a silicon photodiode to construct reflectivity images.
The positions of the alignment marks are obtained from the images, which is used to perform coordinate transformation for keeping the nanobeam devices in the focal plane during automated scans \cite{Ishii2015}.
All measurements are performed at room temperature in a dry nitrogen environment.
Continuous-wave excitation is used unless otherwise noted.

\paragraph*{Device selection and characterization}
To find devices for detailed investigation, we collect PL spectra as we scan along each nanobeam for 10~$\mu$m, constructing colormaps of PL intensity as a function of position and wavelength (Supplementary Fig.~1).
The signature of optical coupling is the narrow peak along the wavelength axis, which corresponds to the high quality factor cavity mode.
To extract such data, we perform two-dimensional peak detection, and then nonlinear curve fitting is used to determine the linewidths for each peak.
We select devices for detailed characterization when the linewidth is below 5~nm, sufficiently narrower than the typical linewidths of nanotubes.
After further screening by PL characterization, less than 20 devices are selected for time-resolved measurements out of 300,000 devices scanned.

\paragraph*{Collection efficiencies}
We use FDTD simulations to obtain the far-field radiation patterns from which collection efficiencies are calculated.
A resonant source is placed to obtain the pattern for the cavity mode, and $C_{\text{cav}}$ is computed from the fraction of photon flux within the numerical aperture of the objective lens.
The nanotube radiation pattern is obtained in a similar manner, but the excitation source is placed at the node of the mode profile and is spectrally detuned by 26 nm.
In addition to the collection efficiency of the objective lens, we also include a correction factor in $C_{\text{CNT}}$ to take into account the emission from the length of the tube beyond the width of the nanobeam.
The additional factor is given by the entire length of the tube normalized with respect to the spatial width of the cavity mode.
The entire tube length is calculated geometrically by the tube angle with respect to the nanobeam axis (Fig.~\ref{Fig2}f inset), and the width of the cavity mode is taken to be 540~nm.

\paragraph*{Time-resolved measurements}
The same excitation laser as the PL characterization is used but the output is switched to $\sim$100~fs laser pulses with a repetition rate of 76~MHz and an average power of 100~nW.
The laser wavelength and polarization are adjusted to maximize the PL intensity.
After spectral rejection of the excitation laser, we couple the emission to a single-mode optical fiber which limits detection to a 1.8-$\mu$m diameter spot at the sample.
A fiber-coupled superconducting single-photon detector with a timing jitter of 32~ps is used to detect the photons, and decay curves are constructed by time-correlated single-photon counting.
Wavelength dependent instrument response function (IRF) is obtained by coupling broadband supercontinuum pulses through a spectrometer with a spectral window of $\sim$1~nm.
We fit the decay curve with a biexponential function convoluted with the IRF to obtain $\tau_1$, and the data in the time range beyond $10\tau_1$ is used to characterize $\tau_2$ by fitting to a single exponential function.
The error for the acceleration factor is computed from the $1\sigma$ confidence intervals of the fits and the standard deviation of the lifetime data in free space.
The shortest $\tau_1$ resolvable by the fitting process is approximately 13~ps. 
Since $\tau_1$ in free space is between 54.6~ps to 64.2~ps for the 4 chiralities in Fig.~\ref{Fig5}, the upper bound of the measurable acceleration factor ranges from 4.2 to 4.9.

\section*{acknowledgments}
This work is supported by MIC (SCOPE 191503001), JSPS (KAKENHI JP20H02558, JP20J00817, JP20K15199, JP19J10319), MEXT (Nanotechnology Platform JPMXP09F19UT0072), and RIKEN (Incentive Research Project). H.M. and D.Y. are supported by JSPS (Research Fellowship for Young Scientists). H.M. acknowledges support from RIKEN (Junior Research Associate Program). FDTD calculations are performed using HOKUSAI BigWaterfall supercomputer at RIKEN. We acknowledge the Advanced Manufacturing Support Team at RIKEN for technical assistance.

\section*{Author Contributions}
Y.K.K. conceived the experiments and supervised the project. H.M. fabricated the devices and performed the measurements. D.Y. assisted CNT growth, device scan measurements, and FDTD simulations. A.I. collected the length dependence data for lifetimes in air-suspended nanotubes. All authors discussed the results and commented on the manuscript.

\section*{Competing financial interests}
The authors declare no competing financial interests.

\end{document}